\documentclass[aps,ams,prb,twocolumn,showpacs]{revtex4}

\usepackage{epsfig}
\usepackage{graphicx}  % Include figure files
\usepackage{dcolumn}   %Align table columns on decimal point
\usepackage{amssymb,amsmath}

\newcommand{\be}{\begin{equation}} \newcommand{\ee}{\end{equation}}
\newcommand{\bea}{\begin{eqnarray}} \newcommand{\eea}{\end{eqnarray}}
\newcommand{\no}{\nonumber}

\begin{document}

\title{Quantum chaos in nanoelectromechanical systems}
\author{Andr\'e Gusso$^1$, M. G. E. da Luz$^1$, and Luis
G. C. Rego$^2$} \affiliation{1-Departamento de F\'{\i}sica,
Universidade Federal do Paran\'a, Curitiba, PR, 81531-990, Brazil, \\
2-Departamento de F\'{\i}sica, Universidade Federal de Santa Catarina,
Florian\'opolis, SC, 88040-900, Brazil}

\date{\today}

\begin{abstract}

We present a theoretical study of the electron-phonon coupling in suspended nanoelectromechanical
systems (NEMS) and investigate the resulting quantum chaotic behavior. The phonons are associated with
the vibrational modes of a suspended rectangular dielectric plate, with free or clamped boundary
conditions, whereas the electrons are confined to a large quantum dot (QD) on the plate's surface. The
deformation potential and piezoelectric interactions are considered. By performing standard
energy-level statistics we demonstrate that the spectral fluctuations exhibit the same distributions as
those of the Gaussian Orthogonal Ensemble (GOE) or the Gaussian Unitary Ensemble (GUE), therefore
evidencing the emergence of quantum chaos. That is verified for a large range of material and geometry
parameters. In particular, the GUE statistics occurs only in the case of a circular QD. It represents
an anomalous phenomenon, previously reported for just a small number of systems, since the problem is
time-reversal invariant. The obtained results are explained through a detailed analysis of the
Hamiltonian matrix structure.
\end{abstract}

\maketitle

\section{Introduction}

The possibility of engineering devices at the nano and micro scales has created the conditions for
testing fundamental aspects of quantum theory \cite{Brandes}, otherwise difficult to probe in natural
atomic size systems. In particular, quantum dots (QD) have largely been considered as a physical
realization of quantum billiards \cite{Berggren,Bruus,Beenakker} and mesoscopic structures have played
an important role in the experimental study of quantum chaos \cite{Stockmann}, mainly through the
investigation of the transport properties of quantum dot \cite{Marcus} and quantum well \cite{Fromhold}
structures in the presence of magnetic field. However, some extraneous effects can prevent the full
observation of the quantum chaotic behavior. For instance, impurities and soft confining potentials may
mask the chaotic dynamics predicted for some semiconductor quantum billiards ({\it e.g}., the
stadium)\cite{Berggren} and the incoherent influence of the bulk on the electronic dynamics hinders the
observation of the so called eigenstate scars \cite{Heller} in quantum corrals \cite{Crommie}.
Furthermore, Random Matrix Theory (RMT) predictions for the Coulomb blockade peaks in quantum dots may
fail as a result of the coupling with the environment \cite{Madger}.

Alternatively, suspended nanostructures are ideal candidates for implementing and investigating
coherent phenomena in semiconductor devices, because, at low temperatures, they provide excellent
isolation for the quantum system from the bulk of the sample \cite{nano,Blencowe}. The
nanoelectromechanical systems (NEMS), in particular, are specially suited to study the effects of a
phonon bath on the electronic states, possibly leading to a chaotic behavior. Such point is of
practical relevance since it bears the question of the stability of quantum devices
\cite{Chuang,Cleland}, whose actual implementation could be prevented by the emergence of chaos
\cite{Georgeot}.

In a recent paper \cite{rego}, we have shown that in fact suspended nanostructures can display quantum
chaotic behavior. In this article we extend such studies and perform a detailed analysis of the
coupling between the phonons of a suspended nanoscopic dielectric plate and the electrons of a
two-dimensional electron gas (2DEG). The phonons are associated with the vibrational modes of a
suspended rectangular plate ({\it i.e.}, the phonon cavity) and the 2DEG (in the free electron
approximation) is confined to a large quantum dot (billiard) built on the plate's surface.

\begin{figure}[h]
\includegraphics[width=8cm]{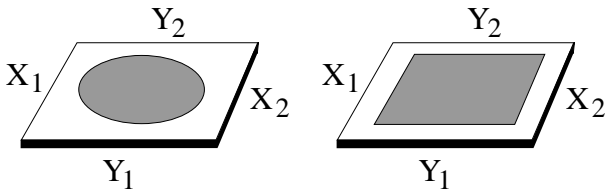}
\caption{Schematics of the suspended nanoelectromechanical structures depicting the cases of a circular
and a rectangular quantum dot on the surface of a suspended dielectric plate.} \label{system}
\end{figure}

Two different scenarios are considered for the shape of the quantum dot: circular and rectangular
geometries (Fig. \ref{system}), which yield distinct chaotic features. As for the coupling mechanisms,
we take into account the deformation and piezoelectric potentials. By performing energy-level
statistics we show that, for sufficiently strong electron-phonon coupling, such electromechanical
nanostructures can exhibit quantum chaos for a large range of material and geometry parameters. The
resulting spectral correlation functions, which depend on the geometry and location of the center of
the QD on the surface of the plate as well as on the plate's boundary conditions (free or clamped), are
those expected from the Gaussian Orthogonal Ensemble (GOE) or Gaussian Unitary Ensemble (GUE) of the
Random Matrix Theory (RMT) \cite{Mehta}. We present a detailed explanation for the occurrence of such
different statistics distributions. Noteworthy are the results for the circular QD, since in this case
the GUE statistics can be obtained in spite of the fact that the system is time-reversal invariant. By
investigating the influence of material and geometrical parameters on the unfolding of chaos, we
indicate the conditions for its experimental observation.

%%%%%%%%%%%%%%%%%%%%%%%%%%%%%%%%%%%%%%%%%%%%%%%%
\section{The system Hamiltonian}
%%%%%%%%%%%%%%%%%%%%%%%%%%%%%%%%%%%%%%%%%%%%%%%%

The full Hamiltonian of the problem is composed of three parts: the phonons, the electrons and the
electron-phonon interactions, which are formulated in the sequence.

%%%%%%%%%%%%%%%%%%%%%%%%%%%%%%%%%%%%%%%%%%%%%%%%
\subsection{Phonons}
%%%%%%%%%%%%%%%%%%%%%%%%%%%%%%%%%%%%%%%%%%%%%%%%

At temperatures below 1 Kelvin the acoustic phonon mean free path in SiN (silicon nitride), for
instance, can be as large as 10 $\mu$m \cite{phonongas}. This implies that a plane wave acoustic phonon
propagating through a suspended mesoscopic system whose dimensions are much smaller than this mean free
path hits the boundaries many times during its expected lifetime, giving rise to standing waves. Thus,
the phonons in such systems can be described in terms of the plate's normal modes of deflection,
instead of the plane wave phonon description that is more appropriate for bulk systems. Therefore, in
the following we associate the phonons with the vibrational modes of the suspended mesoscopic system.
In addition, at low temperatures the semiconductor can be treated as a continuum elastic material due
to the large wavelength of the phonons.

To obtain the long wavelength vibrational modes of the plate we use the Classical Plate Theory (CPT)
approximation \cite{Graff}. The CTP describes adequately the vibrations of a plate whose thickness is
much smaller than its lateral dimensions, which is the characteristic of our NEMS. The deflections of a
plate lying in the $(x,y)$ plane are thus described by a vector field $\left[U({\bf r}) \, \hat{\imath}
+ V({\bf r}) \, \hat{\jmath} + W({\bf  r}) \, \hat{k}\right] \exp(- i \omega t)$ of components
\begin{eqnarray}
& & U(x,y,z) = - z \frac{\partial W}{\partial x}, \ \
V(x,y,z) = - z \frac{\partial W}{\partial y}\ , \\
\label{W} & & W(x,y) =  \sum_{m,n} A_{mn} X_m(x)Y_n(y).
\label{w(x,y)}
\end{eqnarray}
In Eq. (\ref{w(x,y)}), $W(x,y)$ is written in terms of the one-dimensional transverse modes $X_m$ and
$Y_n$, which are the solutions of the Bernoulli-Euler equation \cite{Graff,Leissa} under the
appropriate boundary conditions. Considering that each of the four sides of the plate can be either
clamped (C) or free (F) (corresponding to the Dirichlet or Neumann boundary conditions, respectively),
we have
\bea
X_m(x) &=& \sin[k_mx] \pm \sinh[k_mx] \no \\
       & &  + \zeta \left\{\cos[k_mx] \pm \cosh[k_mx]\right\} \ ,
\label{b-e}
\eea
where
\be
\zeta = \frac{\cos[k_m L_x] - \cosh[k_m L_x]} {\sin[k_m L_x] +
\sinh[k_m L_x]}.
\ee
Likewise for $Y_n(y)$.
The signs in Eq. (\ref{b-e}) are positive (negative) for the FF (CC
and CF) boundary conditions \cite{orthonormal}.
The $k_m$'s are solutions of $\cos[k L_x] \cosh[k L_x] = 1$.

Under given boundary conditions, the Rayleigh-Ritz method is used to
obtain the coefficients $A_{mn}^\alpha$ of Eq. (\ref{w(x,y)}) and the
eigenfrequencies $\omega_\alpha$ corresponding to the eigenmode
${\bf u}_\alpha({\bf r})$ of the plate.
It is done by imposing $\partial\mathcal{U}/\partial A_{mn} = 0$
to the energy functional ${\cal{U}} = \int dx dy \, [{\cal{K}}(x,y) -
{\cal{V}}(x,y)]$ \cite{Leissa}, where the kinetic and strain energies
are written as
\begin{eqnarray}
{\cal{K}}(x,y) &=& \rho_{_{2D}} \frac{\omega^2}{2} W^2(x,y), \\
\label{T_xy}
{\cal{V}}(x,y) &=&
\frac{D}{2} \left(\frac{\partial^2 W}{\partial x^2} +
\frac{\partial^2W} {\partial y^2} \right) - (1-\nu) D
\nonumber \\
& & \times \left( \frac{\partial^2 W}{\partial x^2}
\frac{\partial^2 W}{\partial y^2} -
\left(\frac{\partial^2 W}{\partial x \partial y}\right)^2 \right).
\label{V_xy}
\end{eqnarray}
Here $\rho_{_{2D}}$ denotes the two-dimensional density, $\nu$ is the Poisson constant and $D$ is the
rigidity constant.

\begin{figure}[h]
\includegraphics[width=7cm]{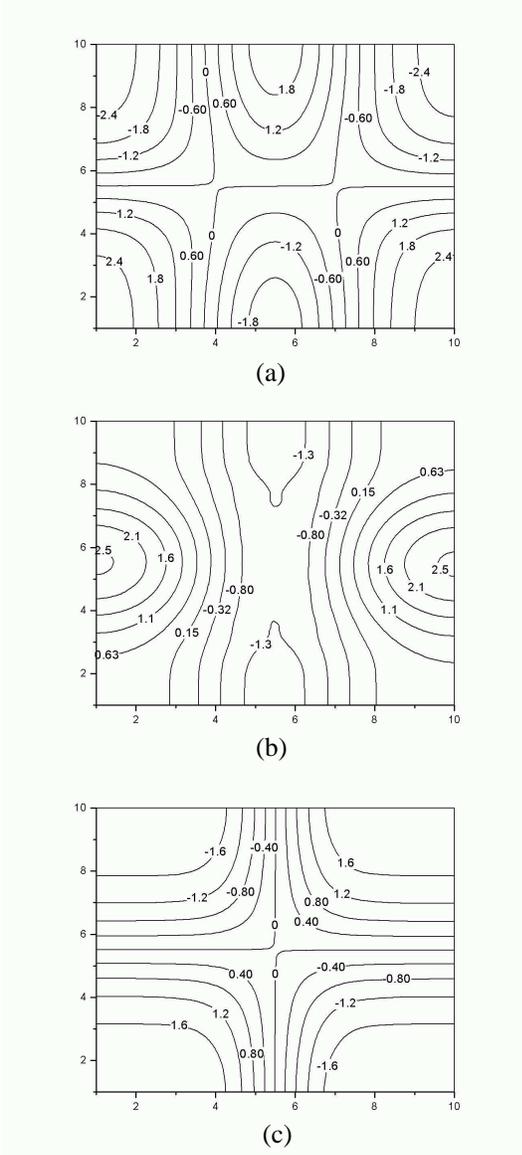}
\caption{Contour plots for the cavity deflection mode SA2 (the second symmetric/antisymmetric
eigenmode) for the \{FFFF\} boundary conditions, calculated at the surface of the plate: (a) $W(x,y)$,
(b) $U(x,y)$, and (c) $V(x,y)$. The amplitudes and lateral dimensions are given in arbitrary units.}
\label{modes}
\end{figure}

In the CPT approximation, the most important motion is that in the $z$ direction, given by $W(x,y)$,
whereas the displacements along the $x$ and $y$ directions are described approximately by the first
term of an orthogonal basis expansion. Therefore, an arbitrary transverse motion can be expanded in the
basis of the orthonormal vibrational modes $W_\alpha(x,y)$, for which \bea & & \langle W_{\alpha} (x,y)
| W_{\beta} (x,y) \rangle = \delta_{\alpha \beta}, \no \\ & & \sum_{\alpha}  W_{\alpha} (x,y)
W_{\alpha} (x',y') = \delta(x-x') \delta (y-y'). \label{ortho} \eea The $U_\alpha$ and $V_\alpha$
components are only approximately orthonormal. As an illustration, we show in Fig. \ref{modes} the
vector field components $(U_\alpha,V_\alpha, W_{\alpha})$ of the SA2 eigenmode (i.e., the second
symmetric/antisymmetric eigenmode) for the \{FFFF\} boundary conditions. Hereafter, we refer to the set
of boundary conditions of the plate, either C or F, as \{$X_1 X_2 Y_1 Y_2$\}, in accordance to Fig. 1.

An arbitrary vibration field of the cavity is written in terms of its deflection modes $\alpha$ as \bea
{\bf u}({\bf r},t) &=& \sum_{\alpha} \left[Q_{\alpha}(t) + Q^*_{\alpha}(t)\right]
\nonumber \\
& & \times \left[ U_{\alpha}({\bf r}) \, \hat{\imath} + V_{\alpha}({\bf r}) \, \hat{\jmath} +
W_{\alpha}({\bf r}) \, \hat{k} \right], \label{modo_geral} \eea together with the normal coordinates
$Q_{\alpha}(t) = Q_{\alpha} \exp[-i\omega_\alpha t]$. In writing Eq. (\ref{modo_geral}), we have taken
into account that the modes $\alpha$ are real.

To provide the same level of description for the elastic and electronic degrees of freedom of the
electromechanical nanostructure, we perform the canonical quantization of the vibration field given by
Eq. (\ref{modo_geral}). As a result, we associate the classical field ${\bf u}({\bf r},t)$ with the
quantum operator $\hat{{\bf u}}({\bf r},t)$, which must satisfy the equal-time commutation relation
$[{\bf \hat{u}}_j({\bf r},t), {\bf \hat{\pi}}_j({\bf r'},t)] = i \hbar \delta({\bf r - r'})$, with the
conjugate momentum operator ${\bf \hat{\pi}} ({\bf r},t) = m \partial_t{\bf\hat{u}}$. Particularly, for
the $\hat{k}$ component of the field we have \begin{eqnarray}[W(x,y),\pi_z(x,y)] = i \hbar \delta (x -
x) \delta (y - y). \label{comut_w} \end{eqnarray} But if we write ($\mathcal{X}_\alpha(t) \equiv \left[
\hat{Q}_\alpha(t) + \hat{Q}_\alpha^\dagger \right]$)
\begin{eqnarray}
W(x,y,t) &=& \sum_\alpha \mathcal{X}_\alpha(t) W_\alpha,
\nonumber \\
\pi_z(x,y,t) &=& \rho V \sum_\alpha (-i \omega_\alpha) \mathcal{X}_\alpha(t) W_\alpha = \sum_\alpha
\mathcal{P}_\alpha(t) W_\alpha,
\nonumber \\
\end{eqnarray}
the commutation relation (\ref{comut_w}) yields \be [W(x,y),\pi_z(x,y)] = \sum_{\alpha,\beta}
[\mathcal{X}_\alpha, \mathcal{P}_\beta] W_\alpha W_\beta. \ee Then, by requiring that
$[\mathcal{X}_\alpha,\mathcal{P}_\beta] = i \hbar \delta_{\alpha, \beta}$ one can use Eq. (\ref{ortho})
to show that Eq. (\ref{comut_w}) is satisfied. Thus, $\mathcal{X}_\alpha$ and $\mathcal{P}_\alpha$ are
canonically conjugated operators, satisfying $[\mathcal{X}_\alpha,\mathcal{X}_\beta] =
[\mathcal{P}_\alpha,\mathcal{P}_\beta] = 0$ as well.

The normal coordinates are now the quantum mechanical operators $\hat{Q}_\alpha(t)$ and
$\hat{Q}_\alpha^\dagger(t)$, which are used to define the dimensionless number operators \be
a^\dagger_\alpha = \sqrt{\frac{2V \rho \omega_\alpha}{\hbar}} \hat{Q}^\dagger_\alpha, \qquad a_\alpha =
\sqrt{\frac{2V \rho \omega_\alpha}{\hbar}} \hat{Q}_\alpha. \ee From the previous commutation relations
it can be shown that $[a_\alpha (t),a_\beta^\dagger (t)] = \delta_{\alpha, \beta}$ and
$[a_\alpha,a_{\alpha'}]=[a^\dagger_\alpha,a^\dagger_{\alpha'}]=0$. Therefore, $a^\dagger_\alpha$ and
$a_\alpha$ are the creation and annihilation operators of the phonon deflection modes $\alpha$ and the
vibration field operator ${\bf \hat{u}}({\bf r},t)$ is
\begin{equation}
\hat{{\bf u}} = \sum_{\alpha} \frac{[a_{\alpha}(t) +
a^\dag_{\alpha}(t)]} {\sqrt{2\ V\rho\ \omega_{\alpha}/\hbar}}
\left[U_{\alpha}({\bf r})\, \hat{\imath} + V_{\alpha}({\bf r})\,
\hat{\jmath} + W_{\alpha}({\bf r})\, \hat{k} \right].
\label{modo_quantico}
\end{equation}

%%%%%%%%%%%%%%%%%%%%%%%%%%%%%%%%%%%%%%%%%%%%%%%
\subsection{Electrons}
\label{el-gas}
%%%%%%%%%%%%%%%%%%%%%%%%%%%%%%%%%%%%%%%%%%%%%%%

We consider the free electron approximation and assume the electrons to be completely confined to a
narrow quantum dot, forming a quasi-2DEG of thickness $d$. The normalized electronic eigenstates are
written as $\phi_{\kappa,\gamma}({\bf r}) = \varphi_{\kappa}(x,y) \sqrt{2/d} \sin[\gamma \pi z/d]$. Due
to the quasi-2D assumption, the electrons always occupy the lowest state in the $z$ direction, so in
our calculations we set the quantum number $\gamma=1$.

For the rectangular QD of sides $L_x$ and $L_y$, we have
\be
\varphi_{\kappa}(x,y) = \frac{2}{\sqrt{L_x L_y}}
\sin[ \frac{p \pi x}{L_x}] \sin[\frac{q \pi y}{L_y}],
\ee
with $\kappa \equiv (p, q)$ and $p, q$ assuming positive integer
values.
The corresponding eigenenergies are
\be
E_{\kappa,\gamma=1} = \frac{\pi^2 \hbar^2}{2 m_e}
\left( \frac{p^2}{L_x^2} + \frac{q^2}{L_y^2} + \frac{1}{d^2} \right),
\label{E_square}
\ee
where $m_e$ is the effective electron mass in the QD.

For the circular QD of radius $R$, we have
\begin{equation}
\varphi_{\kappa}(r,\theta) = \frac{{\mbox J}_{|l|}
\left(\alpha_{l\nu} \frac{r}{R}\right) \exp[i \, l \, \theta]}
{\sqrt{\pi}R|J_{|l|+1}(\alpha_{l\nu})|},
\label{pureelectron}
\end{equation}
with $\kappa \equiv (l, \nu)$, $l = 0, \pm 1, \pm 2,\ldots$ and
$\alpha_{l \nu}$ the $\nu$-th root of the Bessel function of order
$|l|$.
Here, the eigenenergies are
\be
E_{\kappa,\gamma=1} = \frac{\hbar^2}{2 m_e}
\left(\frac{\alpha_{l \nu}^2}{R^2} + \frac{\pi^2}{d^2}\right).
\label{E_circle}
\ee

%%%%%%%%%%%%%%%%%%%%%%%%%%%%%%%%%%%%%%%%%%%%%%%
\subsection{Electron-phonon interactions}
%%%%%%%%%%%%%%%%%%%%%%%%%%%%%%%%%%%%%%%%%%%%%%%

The electrons interact with the lattice vibrations through different mechanisms, depending on the
characteristics of the solid and the temperature. In addition, from the theoretical point of view there
can be several approaches to describe the coupling between electrons and phonons. Next we formulate the
electron-phonon interaction terms that are more relevant to our problem.

\subsubsection{Deformation potential - DP}

At low temperatures only the long wavelength acoustic modes are populated and the semiconductor can be
described by the continuum approximation. As a result of the cavity deflections, local volume changes
take place, thus modifying the lattice constant and the electronic energy bands. In first order, such
volume changes are due to longitudinal (compressional) acoustic modes and the scattering potential
acting on the electrons is proportional to $\hat{\Delta}({\bf r}) = \nabla \cdot \hat{{\bf u}}({\bf
r})$. Therefore, the Hamiltonian for the DP interaction is \bea \hat{H}_{DP} &=& C_{DP}
\int_{\mathcal{D}} d{\bf r} \ \Psi^\dagger({\bf r}) \nabla \cdot  \hat{{\bf u}}({\bf r})
\Psi({\bf r}) \no \\
&=& C_{DP} \sqrt{\frac{\hbar}{2 V \rho}}
\sum_{\alpha \, \kappa'' \kappa'}
\frac{V_{\alpha \, \kappa'' \, \kappa'}^{DP}}{\sqrt{\omega_{\alpha}}}
\, b^\dag_{\kappa''} \left[a_{\alpha}^\dag + a_{\alpha}\right]
b_{\kappa'},
\nonumber \\
\label{DP} \eea with $C_{PD}$ denoting the deformation potential constant for the material. $\Psi ({\bf
r}) = \sum_{\kappa} b_{\kappa} \phi_{\kappa}({\bf r})$ and $\Psi^{\dagger} ({\bf r}) = \sum_{\kappa}
b^{\dagger}_{\kappa} \phi_{\kappa}^{*}({\bf r})$ are the electron field operators and $b_\kappa$
($b^\dagger_\kappa$) is the fermionic creation (annihilation) operator satisfying usual
anti-commutation relations.

The integral is performed over the volume $\mathcal{D}$ comprising
the 2DEG.
In the above equation, $V_{\alpha \, \kappa'' \, \kappa'}^{DP}$ is
given by
\be
V_{\alpha \, \kappa'' \, \kappa'}^{DP} =
\int_{\mathcal{D}} d{\bf r} \ \phi^*_{\kappa''} \nabla \cdot
\left(U_{\alpha} \hat{\imath} + V_{\alpha} \hat{\jmath} + W_{\alpha}
\hat{k} \right) \phi_{\kappa'}.
\ee
Since
\be
\nabla \cdot \left(U_{\alpha} \hat{\imath} +
V_{\alpha} \hat{\jmath} + W_{\alpha} \hat{k} \right)
= -z \left(\frac{\partial^2 W_\alpha}{\partial x^2} +
\frac{\partial^2 W_\alpha}{\partial y^2}\right),
\ee
we have from Eq. (\ref{w(x,y)}) that
\be
V_{\alpha \, \kappa'' \, \kappa'}^{DP} =
- \sum_{mn} A^{\alpha}_{mn} \int_{\mathcal{D}} d{\bf r} \ z \,
\phi^*_{\kappa''} \left( X''_m Y_n + X_m Y''_n \right)
\phi_{\kappa'}.
\label{F}
\ee

\subsubsection{Piezoelectric potential - PZ}

In piezoelectric materials, the acoustic lattice vibrations produce
polarization fields that act back on the vibrational modes.
The result is a set of coupled equations for the acoustic and
polarization fields.
However, taking into account the difference between the sound
and light velocities, such equations can be decoupled, yielding the
following electric field in the semiconductor \cite{Auld}
\begin{equation}
{\bf E} = -2 \frac{\varrho_{14}}{\epsilon}
(\varepsilon_{yz}, \varepsilon_{xz}, \varepsilon_{xy}).
\label{piezo_field}
\end{equation}
$\varrho_{14}$ and $\epsilon$ are, respectively, elements of the
piezoelectric and  dielectric tensors.
Expression (\ref{piezo_field}) is obtained taking into account the
cubic symmetry of the lattice.
Furthermore, from the CPT approximation, the strain tensor elements
$\varepsilon_{xz}=\varepsilon_{yz}=0$ and
\begin{eqnarray}
\varepsilon_{xy} = \left(\frac{\partial U}{\partial y} +
\frac{\partial U}{\partial x} \right) =
-z \frac{\partial^2 W}{\partial x \partial y}.
\end{eqnarray}
Therefore, for a given transverse mode $\alpha$, the resulting electric
field is perpendicular to the plane of the cavity
\begin{eqnarray}
{\bf E}_{pz} = 2 \Lambda(z) \ \frac{\varrho_{14}}{\epsilon} \
\frac{\partial^2 W_{\alpha}}{\partial x \partial y} \hat{k},
\label{field_pz}
\end{eqnarray}
with $\Lambda(z) = d (2 d - z)/2$.

The potential energy of the electrons can be written as
$-e \int {\bf E}_{pz} \cdot d{\bf l}$, leading to
\begin{eqnarray}
2\frac{e \varrho_{14}}{\epsilon} \Lambda(z) \sum_{\alpha}
\sqrt{\frac{\hbar}{2V\rho\omega_{\alpha}}} \,
[a_{\alpha} + a^\dag_{\alpha}] \frac{\partial^2
W_{\alpha}}{\partial x \partial y}.
\end{eqnarray}
Finally, we write down the PZ electron-phonon Hamiltonian as ($C_{PZ} =   2 e \varrho_{14}/\epsilon$)
\begin{equation}
\hat{H}_{PZ} = C_{PZ} \sqrt{\frac{\hbar}{2 V \rho}}
\sum_{\alpha \, \kappa'' \, \kappa'}
\frac{V_{\alpha \, \kappa'' \, \kappa'}^{PZ}}
{\sqrt{\omega_{\alpha}}} \, b^\dag_{\kappa''}
\left[a_{\alpha}^\dag + a_{\alpha}\right] b_{\kappa'},
\label{G}
\end{equation}
with
\begin{equation} V_{\alpha \, \kappa'' \, \kappa'}^{PZ} = \sum_{mn}
A^{\alpha}_{mn} \int_{\mathcal{D}} d{\bf r} \, \Lambda(z) \,
\phi^*_{\kappa''} X'_m Y'_n \, \phi_{\kappa'}.
\label{GV}
\end{equation}

\subsection{The full Hamiltonian}

The total Hamiltonian of the system, when both the DP and PZ
interactions are included, is
\bea
\hat{H} &=& \hat{H}_{el} + \hat{H}_{ph} + \hat{H}_{el-ph}
\nonumber \\
&=& \sum_\kappa E_\kappa b^\dagger_\kappa b_\kappa +
\sum_\alpha \left( \hat{n}_\alpha +\frac12 \right) \hbar \omega_\alpha
+ \hat{H}_{DP} + \hat{H}_{PZ}.
\no \\
\eea

The basis in which $\hat{H}$ is represented is constructed as the product of the one-electron state
$|\phi_\kappa \rangle$ with the multi-phonon state $|n_1,n_2,n_3,...,n_N \rangle$. Here, $n_\alpha =
0,1, \ldots, n$ denotes the number of phonon quanta in mode $\alpha$, with maximum population set by
$n$. A total of $N$ distinct phonon modes are considered. The values of $n$ and $N$ are chosen to be
compatible with the thermodynamics of the system. At low temperatures (below 1K) $n$ is of the order of
a few tens, the average phonon occupation number. On the other hand, $N$ ranges from $\sim 10$, at the
lowest temperatures, up to $\sim 30$ at the highest ones. Hence, in the numerical calculations we set
$n = 20$ and $N=21$. It has been verified, however, that by varying $n$ and $N$ through a considerably
wide range does not alter our main results.

A typical basis vector is written, for a given ${\bf n} \equiv (n_1,n_2,\ldots,n_N)$, as
\begin{equation}
\left|\kappa; {\bf n} \right> =
|\phi_{\kappa} \rangle \otimes \prod_{\alpha=1}^N
\frac{1}{\sqrt{n_{\alpha}!}} \, (a_{\alpha}^\dag)^{n_{\alpha}} \,
|0\rangle.
\label{base}
\end{equation}
For the diagonalization procedure, we energy-sort the basis set up to a maximum energy value. The
diagonalization is then performed with such set of vectors. Obviously, the energy of each basis state
is given by the sum of the electron and the phonon energies, $E_{\kappa \, {\bf n}} = E_\kappa +
\sum_{\alpha} (n_{\alpha} + 1/2)\hbar \omega_\alpha$. $E_\kappa$ comes from either Eq. (\ref{E_square})
or Eq. (\ref{E_circle}), depending on the specific geometry of the 2DEG. For the formation of the
original basis set $10^5$ levels were taken into account, however, the diagonalization of $\hat{H}$ is
performed in the truncated basis that varied from $3 \times 10^3$ to $15 \times 10^3$ basis states. It
is important to notice that the proportion of different phonon states $|{\bf n}\rangle$ to electron
states $|\phi_{\kappa} \rangle$, comprising the truncated basis, ranges from several tens to about a
hundred, depending on the details of the NEMS. That is, the number of phonon states taking part in the
calculations is much larger.

%%%%%%%%%%%%%%%%%%%%%%%%%%%%%%%%%%%%%%%%%%%%%%%
\section{Energy level statistics}
%%%%%%%%%%%%%%%%%%%%%%%%%%%%%%%%%%%%%%%%%%%%%%%

To determine whether the electron-phonon interaction generates chaos in the NEMS, we consider the
standard approach of looking into the statistical properties of the system eigenenergies
\cite{Stockmann, Gutzwiller}. For completeness, here we give a brief summary of the main ideas. A
general technical overview can be found in Ref. [\onlinecite{Guhr}], whereas a very instructive
discussion is presented for a particular case in Ref. [\onlinecite{Miltenburg}].

Consider the ordered sequence $\{E_1, E_2, \ldots\}$ of eigenenergies
of an arbitrary quantum mechanical problem.
The cumulative spectral function, counting the number of levels
with energy up to $E$, is written as
\be
\eta(E) = \sum_n \Theta(E-E_n).
\ee
In principle, we can always separate $\eta(E)$ into smooth (average)
and oscillatory (fluctuating) parts, so that
\be
\eta(E) = \eta_{\mbox{\scriptsize smooth}}(E) +
\eta_{\mbox{\scriptsize osc}}(E).
\ee
The smooth part is given by the cumulative mean level density
\cite{Guhr}.

To make the analysis independent of the particular scales of the spectrum, one can use the so called
``unfolding'' procedure \cite{Gutzwiller}. It allows the comparison of the results obtained from any
specific system with the predictions of the RMT \cite{Mehta}. The unfolding is  done basically by
mapping the sequence $\{E_1,E_2,\ldots\}$ onto the numbers $\{s_1,s_2, \ldots\}$, where \be s_n \equiv
\eta_{\mbox{\scriptsize smooth}}(E_n). \ee In the new variables, the cumulative spectral function
simply reads $\tilde{\eta}(s) = s + \tilde{\eta}_{\mbox{\scriptsize osc}}(s)$, so that the smooth part
of $\tilde{\eta}$ has unity derivative. Hence, for our statistical studies we consider the resulting
sets $\{s_1,s_2, \ldots\}$.

In this work we calculate two of the most used spectral distributions \cite{Vessen-Xavier}: the
nearest-neighbor spacing distribution, $P(s)$, and the spectral rigidity $\overline{\Delta}_3(l)$. The
$P(s)$ distribution probes the short scale fluctuations of the spectrum. It corresponds to the
probability density of two neighboring unfolded levels $s_n$ and $s_{n+1}$ being a distance $s$ apart.
$\overline{\Delta}_3(l)$ is an example of a distribution that quantifies the long scale correlations of
the energy spectrum. It measures the deviation of the cumulative number of states (within an unfolded
energy interval $l$) from a straight line. Formally \be \overline{\Delta}_3(l) = \frac{1}{l} \left
\langle \mbox{min}_{\{A,B\}} \int_{s_0}^{s_0 + l} ds \left[\tilde{\eta}(s) - A s -B\right]^2 \right
\rangle \ , \ee where $\langle \cdot \rangle$ denotes the averaging over different possible positions
$s_0$ along the $s$ axes. The parameters $A$ and $B$ are chosen to minimize $\left[\tilde{\eta}(s) - A
s -B\right]^2$ in each corresponding interval.

The RMT predicts three different classes of Gaussian ensembles \cite{Mehta,Guhr}, having distinct
$P(s)$ and $\overline{\Delta}_3(l)$: the {\it Gaussian Orthogonal Ensemble} (GOE), the {\it Gaussian
Unitary Ensemble} (GUE), and the {\it Gaussian Sympletic Ensemble} (GSE), constituted by matrices whose
elements are random and obey certain Gaussian-like distribution relations \cite{Stockmann,Mehta}.
Furthermore, these ensembles are invariant under orthogonal, unitary and sympletic transformations,
respectively. Bohigas et al. \cite{Bohigas} conjectured that the spectrum fluctuations of any quantum
chaotic system should have the same features of one of such three cases. This proposal has been firmly
established by theoretical and experimental examinations \cite{Stockmann,Gutzwiller,Guhr}. When spin is
not involved, it is expected that the spectrum statistics of a chaotic system is similar to that
obtained from the GOE (GUE) if it is (is not) time reversal invariant (TRI). However, there are
exceptions to this rule, consisting of a special class of TRI systems with point group irreducible
representations, which does exhibit the GUE statistics \cite{Leyvraz,Keating}. Until recently
\cite{rego}, the only family of systems known to show this anomalous behavior was formed by billiards
having threefold symmetry, implemented experimentally in classical microwave cavities
\cite{Dembowski1,Dembowski2,Schafer}.

For regular (integrable) systems the resulting statistics follow
Poisson $P(s) = \exp[-s]$ and linear
$\overline{\Delta}_3(l)  = l/15$ distributions \cite{Gutzwiller}.
For GOE and GUE, $P(s)$ is described with high accuracy by the
Wigner distributions \cite{Guhr}
\begin{eqnarray}
P(s) &=& \frac{\pi}{2} s \exp[-\frac{\pi}{4} s^2] \ \
\ \qquad (\mbox{GOE}), \nonumber \\
P(s) &=& \frac{32}{\pi^2} s^2 \exp[-\frac{4}{\pi} s^2]
\qquad \, (\mbox{GUE}).
\end{eqnarray}
Finally, $\overline{\Delta}_3(l)$ can be approximated
by the expressions
\begin{eqnarray}
\overline{\Delta}_3(l) &=& \frac{1}{\pi^2}
\left(\ln[2 \pi l] + \gamma -\frac{5}{4} -\frac{\pi^2}{8} \right)
\ \ \ (\mbox{GOE}), \nonumber \\
\overline{\Delta}_3(l) &=& \frac{1}{2\pi^2} \left( \ln[2\pi l] +
\gamma -\frac{5}{4} \right) \qquad \ \ \ (\mbox{GUE}).
\end{eqnarray}
Here, $\gamma = 0.5772\ldots$ is the Euler constant.

To characterize our nanostructures, we compare the numerically calculated distributions $P(s)$ and
$\overline{\Delta}_3(l)$ with the corresponding analytical expressions for the regular and chaotic
cases. Very good statistics are obtained using 2000 up to 2500 energy levels.

%%%%%%%%%%%%%%%%%%%%%%%%%%%%%%%%%%%%%%%%%%%%%%%%
\section{Results}
\label{result}
%%%%%%%%%%%%%%%%%%%%%%%%%%%%%%%%%%%%%%%%%%%%%%%%

\begin{figure}[h]
\includegraphics[width=6.5cm,height=6.5cm]{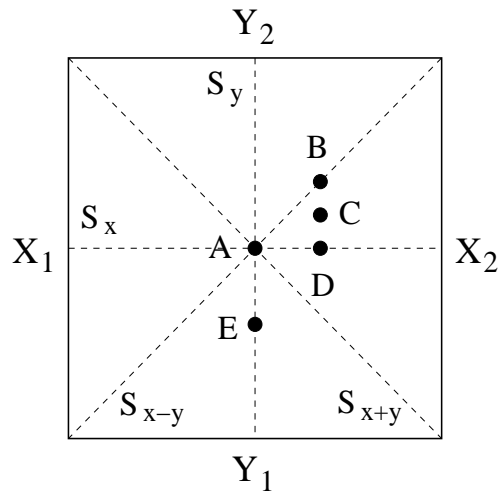}
\caption{The distinct positions, A, B, C, D, and E (on the plate) used as centers for the quantum dot.
The dashed lines represent the possible symmetry axes for the plate phonon modes $\alpha$, depending on
the boundary conditions \{$X_1 X_2 Y_1 Y_2$\}.} \label{abcde}
\end{figure}

We have applied the previous analysis to the eigenenergies of our suspended NEMS, considering a wide
range of material and geometrical parameters, and it was found that chaos emerges in the system for a
sufficiently strong electron-phonon (el-ph) coupling. Although the phenomenon proved to be quite robust
with respect to variations of physical dimensions, boundary conditions and basis size, it was observed
that the chaotic features depend on some material parameters, like the electronic effective mass and
the el-ph coupling constants. The materials used to model the NEMS comprise an AlAs dielectric phonon
cavity and an Al$_{0.5}$Ga$_{0.5}$As quantum dot, where the 2DEG is formed. This choice takes advantage
of the very small lattice parameter mismatch in the interface as well as the large electronic effective
mass of the $X$ valley in AlGaAs  \cite{Adachi}.

In our investigation we varied the DP and PZ interaction strengths (by means of the multiplicative
factors $\beta_{DP}$ and $\beta_{PZ}$), the stiffness tensor elements $c_{11}, c_{12}$ and $c_{44}$,
the mass density of the cavity and the in-plane electron effective mass. As for the geometrical
parameters, we also varied the size and aspect ratio of the dielectric plate, the area of the QD, the
thicknesses of the plate ($\delta$) and of the 2DEG ($d$). More interestingly, however, we considered
different positions for the center of the QD (shown in Fig. \ref{abcde}), which produces distinct
chaotic behaviors.

The most representative results will be presented throughout this section. A detailed analysis is found
in Section \ref{interpreting}.

%%%%%%%%%%%%%%%%%%%%%%%%%%%%%%%%%%%%%%%%%%%%%%%%
\subsection{Circular 2DEG}
\label{circular}

%%%%%%%%%%%%%%%%%%%%%%%%%%%%%%%%%%%%%%%%%%%%%%%%

Here we present a detailed analysis for the spectral statistics of the NEMS containing a circular
quantum dot. Unless  mentioned otherwise, the system comprises  a QD of radius $R=450\ nm$ and
thickness $d = \delta/5$ on the surface of a square phonon cavity of sides $L = 1\ \mu$m and width
$\delta = 40$ nm.

\begin{figure}[h]
\includegraphics[width=8.5cm]{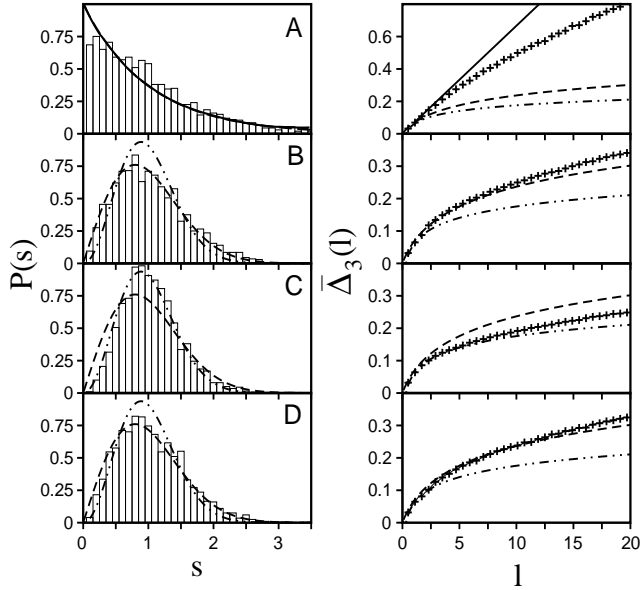}
\caption{ Energy-level statistics for the nanostructure with only the DP interaction, for
$\beta_{DP}=10$ and four different positions (A,B,C, and D) for the center of the QD (refer to Fig.
\ref{abcde}). The cavity boundary conditions are \{FFFF\}. The symbols $+$ represent the numerically
calculated results. The curves indicate the expected behavior for regular (solid), chaotic GOE-type
(dashed), and chaotic GUE-type (dot-dashed) systems.} \label{statistics}
\end{figure}

Once  enough electron-phonon coupling is assured, regular or chaotic spectral features will emerge
depending on the interplay between the symmetries of the cavity phonon modes and the electronic
wavefunctions. In this respect the boundary conditions of the phonon cavity ({\it i.e.}, the dielectric
plate) and the localization of the circular 2DEG play a crucial role. In order to systematically
investigate this effect we make use of the scheme presented in Fig. \ref{abcde}.
%The el-ph interaction destroys all the geometrical invariances,
%except for the reflection symmetries if the center of the quantum
%dot is located on a symmetry axis of the plate's $\alpha$ modes.
Slight displacements of the QD out of the center of the plate suffice to generate different spectral
features. So, the relative coordinates ($-0.5 < x, y < 0.5$) used in the calculations are: A = (0,0), B
= (0.05,0.05), C = (0.05,0.025), D = (0.05,0), and  E = (0,-0.05). Figure \ref{statistics} shows $P(s)$
and $\overline{\Delta}_3(l)$ for cases A,B,C and D in the \{FFFF\} phonon cavity, taking into account
only the DP interaction, with $\beta_{DP}=10$. For A, the spectral statistics indicates a regular
dynamics, but in B the occurrence of quantum chaos is clear and the level distributions are well
described by the predictions of GOE random matrices. The same occurring for D. The more interesting
case, however, is C, for which the statistics belongs to the GUE class, although the system is
time-reversal invariant. The same behavior is obtained for a phonon cavity with \{CCCC\} boundary
conditions \cite{rego}. The reasons for obtaining GUE statistics in this time-reversal invariant system
will be discussed in Section \ref{interpreting}

\begin{figure}[]
\includegraphics[width=8.5cm]{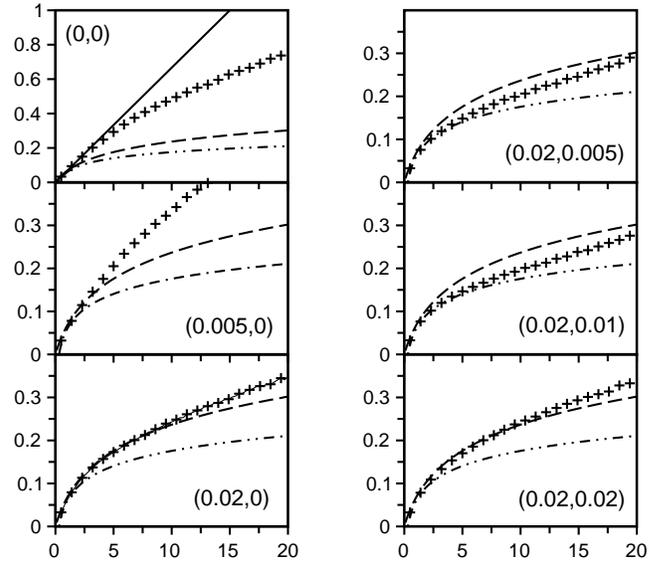}
\caption{ $\overline{\Delta}_3(l) \times l$ for different positions of the center of the circular QD.
The cavity boundary conditions are \{CCCC\} and $\beta_{DP}=10$.} \label{figDP}
\end{figure}

It is also instructive to look at the evolution of the spectral statistics as a function of the
position of the QD. The effect is illustrated by the $\overline{\Delta}_3(l)$ statistics in Fig.
\ref{figDP}, for the \{CCCC\} plate and the parameters of Fig. \ref{statistics}. Leaving A along the
S$_x$ axis the statistics evolves from regular to GOE, at (0.02,0), passing through a mixed behavior at
the locus (0.005,0). Proceeding perpendicularly to the S$_x$ axis, the statistics evolves from GOE
towards GUE, (0.02,0) $\rightarrow$ (0.02,0.005) $\rightarrow$ (0.02,0.01) $\equiv$ C, going back to
GOE when the S$_{x-y}$ axis is reached at (0.02,0.02).

\begin{table*}
\label{tablecirc} \caption{Symmetry axes and spectral statistics for points A,B,C,D and E, as defined
in the text, for a circular 2DEG. Here, 2 GOE means that the statistics can be described by the
uncorrelated superposition of two GOE distributions.}
\begin{tabular}{c c c c c c c}
\hline
\hline
\hspace{0.4 cm} Boundary Conditions \hspace{0.4 cm}  &
\hspace{0.4 cm} Symmetry axes \hspace{0.4 cm}        &
\hspace{0.4 cm} A \hspace{0.4 cm}                    &
\hspace{0.4 cm} B \hspace{0.4 cm}                    &
\hspace{0.4cm}  C \hspace{0.4 cm}                    &
\hspace{0.4 cm} D \hspace{0.4 cm}                    &
\hspace{0.4 cm} E \hspace{0.4cm} \\ \hspace{0.4 cm}  &
\hspace{0.4 cm} for the phonon modes \hspace{0.4 cm} &
\hspace{0.4 cm}                                      &
\hspace{0.4 cm}                                      &
\hspace{0.4 cm}                                      &
\hspace{0.4 cm} & \hspace{0.4cm}
\\ \hline
\{FFFF\}                            &
S$_x$, S$_y$,  S$_{x-y}$, S$_{x+y}$ &
Regular                             &
GOE                                 &
GUE                                 &
GOE                                 &
GOE \\ \{CCCC\}                     &
S$_x$, S$_y$, S$_{x-y}$, S$_{x+y}$  &
Regular   &  GOE  &  GUE  &  GOE  &  GOE \\
\{CCFF\} &  S$_x$, S$_y$                       & 2 GOE     &  GUE  &
GUE  &  GOE  &  GOE \\ \{CFCF\} & S$_{x-y}$
& GOE       &  GOE  &  GUE  &  GUE  &  GUE \\ \{CCCF\} & S$_y$
& GOE       &  GUE  &  GUE  &  GUE  &  GOE \\ \{CFFF\} & S$_x$
& GOE       &  GUE  &  GUE  &  GOE  &  GUE \\ \hline \hline
\end{tabular}
\end{table*}

By the examination of several different scenarios we were able to classify the general behavior of our
system. Table I summarizes the results obtained for the center of the circular 2DEG located at points
A,B,C,D and E with either the DP or the PZ interaction taken into account. Furthermore, we have
considered a comprehensive set of boundary conditions, which are representative of all possible
combinations of the Dirichlet and Neumann conditions for the phonon cavity, thus, producing distinct
symmetry axes for the phonon modes: \{CCCC\}, \{FFFF\}, \{CCFF\}, \{CFCF\}, \{CCCF\}, and \{CFFF\}.
From Table I, we see that the chaotic behavior is determined by the overall (or global) symmetries of
the NEMS, that is, the one that results from the joint combination of the boundary conditions of the
phonon cavity and the position of the QD. For instance, if the boundary conditions are \{CCCC\} and the
QD is located at D, the S$_{y}$, S$_{x-y}$ and S$_{x+y}$ are not symmetry axes for the coupled
electromechanical system. According to Table I, if the present NEMS has: (i) four symmetry axes
(position A for \{CCCC\} and \{FFFF\} plates), the statistics indicates a regular (integrable) problem;
(ii) two symmetry axes (position A for \{CCFF\}), then the statistics corresponds to the uncorrelated
superposition of two distributions of the GOE type \cite{twowigner}; (iii) one symmetry axis, the
results are those of GOE; and finally (iv) no symmetry axes at all, the spectrum exhibits the GUE
statistics.

The boundary conditions determine not only the position of the center of the 2DEG at which regular, GOE
or GUE statistics are obtained, but also the intensity of the quantum chaos. This happens because the
electron-phonon coupling depends on the phonon energies, which vary according to the boundary
conditions. The higher the phonon energy, the stronger the electron-phonon interaction, thus, leading
to spectral fluctuations that are more faithful to the typical chaotic features. The energies of the
phonon modes decrease according to the following sequence: \{FFFF\}, \{CCFF\}, \{CCCC\}, \{CFFF\},
\{CCCF\}, and \{CFCF\}. The energies for the first five boundary conditions are similar, and quantum
chaos can be observed for essentially the same values of the interaction strength $\beta$. For the
\{CFCF\} case, however, the phonon energies can be one order of magnitude smaller than the ones for the
other cases, requiring larger values for the parameters $\beta_{DP}$ and $\beta_{PZ}$ (approximately 3
times larger).

\begin{figure}[h]
\includegraphics[width=6cm]{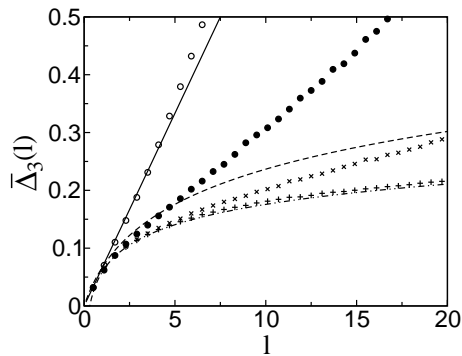}
\caption{The calculated spectral rigidity for various DP coupling constants: $\beta_{DP}$=1 (open
circle), 3 (filled circle), 5 ($\times$) and 10 ($+$). The system corresponds the \{CCCC\} plate with
the QD located at position C. The curves represent the regular (solid), GOE (dashed) and GUE
(dot-dashed) cases.} \label{beta}
\end{figure}

It is, however, important to notice that, regardless of the geometry adopted, the nanostructure shows a
regular spectrum for the {\it bare} parameters of the reference materials. We present in Fig.
\ref{beta} the dependence of the spectral rigidity $\overline{\Delta}_3(l)$ on the electron-phonon
coupling strength. For the locus C of the \{CCCC\} phonon cavity, we take into account only the DP
interaction, with $\beta_{DP}$ = 1, 3, 5, and 10. As $\beta$ increases, the calculated statistics
gradually converges to the GUE prediction. Note that the numerical calculations are never well fitted
by the GOE distribution. The inclusion of more basis states does not alter the observed results. At
this point we observe that the strong el-ph coupling regime ($\beta>3$) can be achieved by using
different materials. For instance, aluminum nitride (AlN) is a strong piezoelectric semiconductor, with
$\varrho_{33}$=1.5\ C/m$^2$, that is currently been used to produce nanomechanical resonators
\cite{AlN}. The piezoelectric constant for GaAs is $\varrho_{14}$=0.16\ C/m$^2$

\begin{figure}[]
\includegraphics[width=6cm]{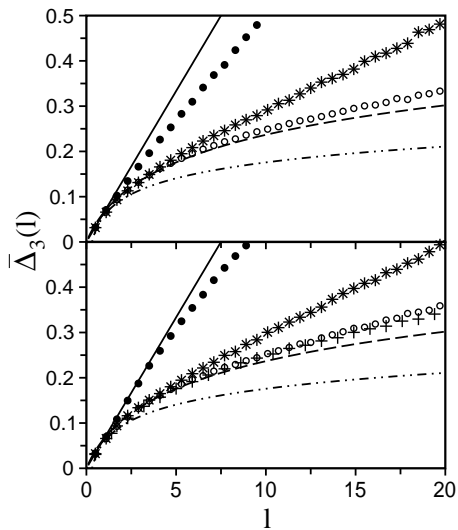}
\caption{Top panel: $\overline{\Delta}_3(l)$ statistics at point A of the \{CFFF\} plate for various
radii of the QD: $R$ = 200 nm (filled dot), 300 nm (star) and 400 nm (open dot); Bottom panel: same
statistics for various $m^*/m_e$ ratios: 0.2 (filled dot), 0.4 (star), 0.6 (open dot), and 0.8 (cross).
$\beta_{DP}$=10.} \label{parameters}
\end{figure}

Next, we summarize the effects of the geometrical and material parameters on the chaotic behavior.
Irrespective of the boundary conditions, when the QD radius $R$ decreases to less than one third of the
plate side $L$, the system starts to become regular and, for about $L/R \approx 5$, the nanostructure
presents no clear signs of chaos in its spectrum. This is illustrated in the top panel of Fig.
\ref{parameters}, by the $\overline{\Delta}_3(l)$ statistics calculated for case A in the \{CFFF\}
plate. On the other hand, an important physical parameter for the occurrence of chaos is the in-plane
electron effective mass $m^*$. It is shown in the bottom panel of Fig. \ref{parameters} that chaos
arises as $m^*/m_e$ is increased. Indeed, for $m^* \gtrsim 0.6$ the system is clearly chaotic, becoming
regular for $m^* \alt 0.2 \, m_e$. As before, similar results hold for other boundary conditions. A
weak dependence on both the density and the value of the stiffness tensors $c_{12}$ and $c_{44}$ is
also observed. In essence, lighter and softer materials favor the appearance of chaos. As for the size
of the dielectric plate, chaos is favored by short and thin plates. The former should be expected
because the phonon energies increase as the area of the plate decreases. However, structures much
smaller than the one here considered do not present a significantly higher tendency to chaotic
behavior.

\begin{figure}
\includegraphics[width=6cm]{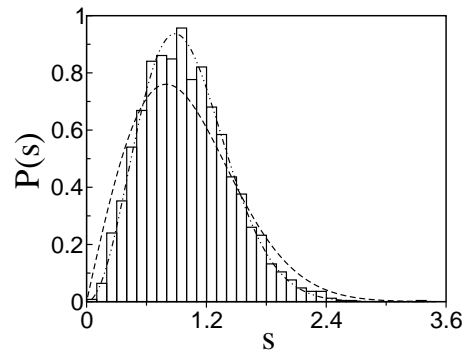}
\caption{The nearest neighbor spacing distribution for the \{CCCC\} plate with the circular QD located
at D. Both the DP and PZ interactions are included  with $\beta_{DP}=\beta_{PZ}=10$. The curves
correspond to GOE (dashed) and GUE (dot-dashed).} \label{DP-PZ}
\end{figure}

So far, we have considered only the DP or PZ interactions acting individually. When acting together,
the spectrum statistics of loci B and D change from GOE to GUE. Fig. \ref{DP-PZ} demonstrates this
effect by showing the $P(s)$ distribution for the circular QD at locus D in the \{CCCC\} plate, for
both the DP and PZ interactions included with $\beta_{DP} = \beta_{PZ} = 10$. The agreement with the
GUE statistics is excellent, in contrast to case D of Fig. \ref{statistics} (we recall that the
\{CCCC\} and \{FFFF\} cases give similar results). Because the AlGaAs alloy is a weak piezoelectric
material, the DP coupling shows a stronger effect in promoting the chaos, whereas the main action of
the PZ interaction (in the presence of DP) is to break the system's overall symmetry. The explanation
for such a change in the spectrum statistics is left to Section V.

%%%%%%%%%%%%%%%%%%%%%%%%%%%%%%%%%%%%%%%%%%%%%%%%
\subsection{Rectangular 2DEG}
\label{rectangular}
%%%%%%%%%%%%%%%%%%%%%%%%%%%%%%%%%%%%%%%%%%%%%%%%

\begin{table*}
\label{tablesquare}
\caption{The same as Table I, but for a rectangular quantum dot.  }
\begin{tabular}{c c c c c c c}
\hline \hline \hspace{0.4 cm} Boundary Conditions \hspace{0.4 cm}    & \hspace{0.4 cm} Symmetry axes
\hspace{0.4 cm}          & \hspace{0.4 cm} A \hspace{0.4 cm}                      & \hspace{0.4 cm} B
\hspace{0.4 cm}                      & \hspace{0.4cm} C \hspace{0.4 cm}                       &
\hspace{0.4 cm} D \hspace{0.4 cm}                      & \hspace{0.4 cm} E \hspace{0.4cm} \\
\hspace{0.4 cm}    & \hspace{0.4 cm} for the phonon modes \hspace{0.4 cm} & \hspace{0.4cm} &
\hspace{0.4 cm} & \hspace{0.4 cm} & \hspace{0.4cm} &
\hspace{0.4 cm}    \\
\hline
\{FFFF\} & S$_x$, S$_y$,  S$_{x-y}$, S$_{x+y}$ & Regular & 2
GOE &  GOE  & 2 GOE & 2 GOE \\
\{CCCC\} & S$_x$, S$_y$,  S$_{x-y}$,
S$_{x+y}$ & Regular & 2 GOE &  GOE  & 2 GOE & 2 GOE \\
\{CCFF\} & S$_x$, S$_y$ & Regular &  GOE  &  GOE  & 2 GOE & 2 GOE \\
\{CFCF\} &    S$_{x-y}$                        & 2 GOE  & 2
GOE &  GOE  &  GOE  &  GOE  \\
\{CCCF\} &  S$_y$ & 2 GOE  &  GOE  &  GOE  &  GOE  & 2 GOE \\
\{CFFF\} &  S$_x$ & 2 GOE  &  GOE  &  GOE  & 2 GOE &  GOE  \\
\hline \hline
\end{tabular}
\end{table*}

\begin{figure}[h]
\includegraphics[width=4.5cm]{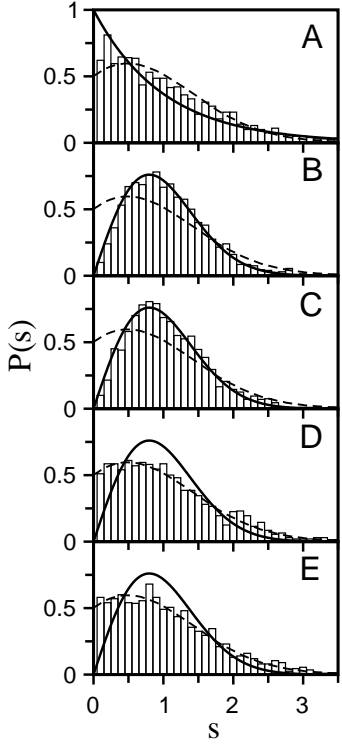}
\caption{$P(s)$ for the \{CCFF\} boundary conditions and the rectangular QD centered at positions A, B,
C, D, and E. Only the DP interaction is considered ($\beta_{DP} = 10$). The dashed lines represent the
uncorrelated superposition of two GOEs and the continuous lines the regular and GOE cases.}
\label{square}
\end{figure}

\begin{figure}[h]
\includegraphics[width=7cm]{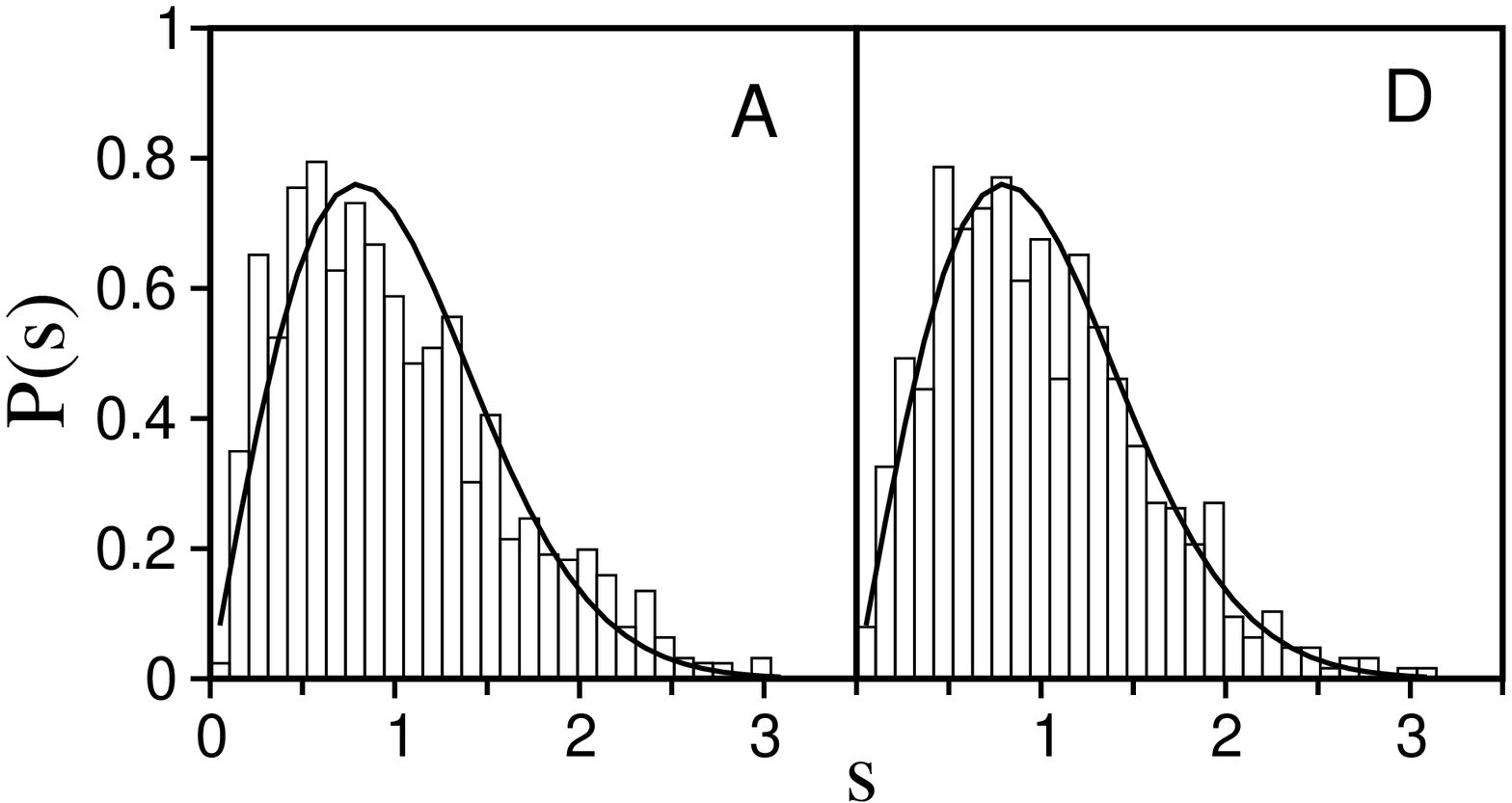}
\caption{$P(s)$ for the \{CFFF\} boundary conditions and both interaction potentials acting together
($\beta_{DP}=\beta_{PZ} = 10$). The rectangular QD is located at positions A (right panel) and D (left
panel).} \label{DPZ}
\end{figure}

Chaos is also observed in the calculations for a rectangular 2DEG interacting with the suspended phonon
cavity. In this section we investigate such nanostructures following the procedures previously
described. The obtained statistics are summarized in Table II for the interactions, either DP or PZ,
taking into account individually. The calculations were made for the same phonon cavity considered
throughout Section \ref{circular}, but now supporting a square QD of sides 400 nm and thickness equal
to the circular case.

Representative results of the $P(s)$ distribution are shown in Fig.\ref{square}, which illustrates the
chaotic behavior of the \{CCFF\} phonon cavity through cases A to E. Here too, the global symmetries of
the system depend on the combination of the symmetry axes of the plate with the position of the square
QD. From extensive simulations we verified that (see Table II) whenever the full problem has only one
global symmetry axis, either S$_x$, S$_y$, S$_{x-y}$ or S$_{x+y}$, the resulting spectral statistics
corresponds to the superposition of two uncorrelated GOEs, contrasting with the case of the circular
QD-NEMS (see Table I). If there are no overall symmetry axes, the statistics is that of the GOE type.
Finally, the spectrum is regular if there are at least two global symmetry axes, namely, the position A
for the \{CCFF\}, \{CCCC\} or \{FFFF\} boundary conditions.

Despite the fact that circular and rectangular QD-NEMSs display chaotic features, it is important to
emphasize that the GUE statistics never occurs for the rectangular 2DEG coupled to a rectangular phonon
cavity. This is, therefore, an effect that results from the interplay between the cylindrical and
rectangular symmetries in the circular QD-NEMS. We discuss this phenomenon in detail in the next
Section.

The dependence of the spectrum statistics on the geometrical and
material parameters is, nonetheless, similar to that observed for the
circular QD. Specifically, heavier in-plane electron effective masses,
lighter and softer materials and larger and thinner quantum dots favor
the appearance of chaos. The \{CFCF\} plate requires stronger
interaction strengths than the other boundary conditions to give rise
to a chaotic spectrum.

Finally, when the system has one global symmetry axis and both the DP and PZ interactions act
simultaneously, the spectrum statistics changes from two uncorrelated GOEs to a single GOE. This effect
is illustrated in Fig \ref{DPZ} for the rectangular 2DEG centered at points A (right panel) and D (left
panel) of the \{CFFF\} cavity. A comparison with Table II evidences the aforementioned transformation.
On the other hand, when the nanostructure displays either regular or chaotic (GOE) distributions for
one of the interactions, the inclusion of the other does not alter the original statistics, regardless
of the strength of the interactions.

%%%%%%%%%%%%%%%%%%%%%%%%%%%%%%%%%%%%%%%%%%%%%%%%%%%%%%%%%%%%%%%%%
\section{Discussion}
\label{interpreting}
%%%%%%%%%%%%%%%%%%%%%%%%%%%%%%%%%%%%%%%%%%%%%%%%%%%%%%%%%%%%%%%%%

It has been shown that distinct geometrical configurations of the QD-NEMS produce different
energy-level statistics, in most cases typical of chaotic dynamics. To understand this effect we
investigate the structure of the Hamiltonian matrix of our systems and explain the previous results in
terms of the underlying symmetries of the problem. At the end, we explain the anomalous GUE statistics
in the light of a more general analysis \cite{Leyvraz,Keating}.

\subsection{The Hamiltonian block structure due to the phonons}
\label{va}

\begin{figure}[h]
\includegraphics[width=8cm]{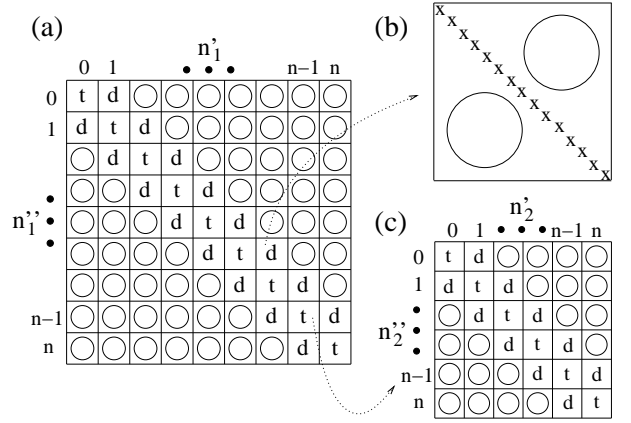}
\caption{ (a) A block of the interaction potential matrix representing fixed electron quantum numbers
and all possible combinations for the phonon states. The small blocks $d$, shown in (b), are diagonal
matrices whose diagonal elements $d_{i i}$ are all equal. The $t$'s forming the main diagonal in (a),
depicted in (c), are also block tridiagonal matrices. Noticeably, d and t in (a) are different from the
corresponding blocks in (c).} \label{blockstructure}
\end{figure}

From the Eqs. (\ref{DP}) and (\ref{G}), for the deformation and piezoelectric potentials, one verifies
that the interaction mechanism is mediated by one phonon processes. This becomes clear by writing the
matrix elements in the basis (\ref{base}) [$\xi \equiv (\kappa,{\bf n})$]
\begin{eqnarray}
h(\xi'';\xi') &=& C \sqrt{\frac{\hbar}{2 \rho V}}
\sum_{\alpha} \frac{{\cal I}_{\alpha \, \kappa'' \kappa'}}
{\sqrt{\omega_{\alpha}}} \, \nonumber \\
& & \times
\left[\delta_{n_1'' \, n_1'} \, \ldots \, \delta_{n_{\alpha}'' \,
n_{\alpha}'-1} \, \ldots \delta_{n_N'' \, n_N'} \right. \nonumber \\
& & + \left.  \delta_{n_1'' \, n_1'} \, \ldots \, \delta_{n_{\alpha}''
\, n_{\alpha}'+1} \, \ldots \delta_{n_N'' \, n_N'} \right],
\label{phononmatrix}
\end{eqnarray}
where $C$ and ${\cal I}_{\alpha \, \kappa'' \kappa'}$ denote, respectively, the appropriate coupling
constant and the overlap of the phonon mode $\alpha$ with the electronic eigenfunctions $\kappa''$ and
$\kappa'$. Notice that the Kronecker $\delta$'s allow only a single phonon transition.

In this representation we have a very particular form for the interaction matrix. Consider the block
${\bf n} \times {\bf n}$, schematically depicted in Fig. \ref{blockstructure} (a). It corresponds to
fixed values for the electron quantum numbers $\kappa''$ and $\kappa'$, but embraces all possible
configurations for the phonon states. The structure of the block ${\bf n} \times {\bf n}$ is such that
the outermost block spans all possible states for the phonon quantum number $n_1 = 0, 1, \ldots,n$.
Then, the next internal block spans the quantum number $n_2$, followed by the inner blocks $n_3,n_4 \,
\ldots \, n_N$. Due to the action of the Kronecker $\delta$'s in (\ref{phononmatrix}), the ${\bf n}
\times {\bf n}$ matrix is block tridiagonal. Consequently, the small blocks $d$ (refer to Fig.
\ref{blockstructure} (b)) must be diagonal, since the interactions are mediated by one phonon only. On
the other hand, the small blocks $t$ are also tridiagonal (Fig. \ref{blockstructure} (c)). Such
self-similar arrangement goes on at all block levels $n_1 \, n_2 \, \ldots \, n_N$.

\subsection{Phonon mode parities}

The reflection symmetries of the phonon cavity lead to phonon modes of well defined parity. Such
properties are examined in this section; for guidance refer to Fig. \ref{abcde}. For instance, when the
boundary conditions at $Y_1$ and $Y_2$ are equal, {\it i.e.}, both C or both F, the modes $\alpha$ have
either a symmetric ($+$) or an anti-symmetric ($-$) parity with respect to S$_x$. The same holds for
S$_y$ regarding the edges $X_1$ and $X_2$. If the boundary conditions are opposite at $X_1$ and $X_2$
and also at $Y_1$ and $Y_2$, then one of the main diagonals of the plate, S$_{x-y}$ or S$_{x+y}$, is
the only symmetry axis. Therefore the phonon modes $\alpha$ have well defined $+$ and $-$ parities
about it. Finally, the cases \{CCCC\} and \{FFFF\} have definite parities about all the symmetry axes:
S$_x$, S$_{y}$, S$_{x-y}$ and S$_{x+y}$.

\subsection{Circular quantum dot}

In the following we analyze the element ${\cal I}_{\alpha \, \kappa''\kappa'}$, appearing in Eq.
(\ref{phononmatrix}). For the circular quantum dot case we have, generalizing Eqs. (\ref{F}) and
(\ref{GV}),
\begin{eqnarray}
{\cal I}_{\alpha\,\kappa''\kappa'} &=& f(d)
\int_{\mathcal{D}_{x y}} dx \, dy \ \ g_{\kappa''\kappa'}(r) \ \
F_\alpha(x,y) \nonumber \\
&\times& \big\{\cos[(l'' - l') \, \theta] + i
\sin[(l'' - l') \, \theta] \big\}.
\label{integralcircle}
\end{eqnarray}
Here, $g_{\kappa''\kappa'}(r)$ denotes the product of Bessel functions coming from Eq.
(\ref{pureelectron}), $r=\sqrt{(x-x_0)^2 + (y-y_0)^2}$ with $(x_0,y_0)$ as the center of the QD,
$\theta$ is measured from the S$_x$ axis and $f(d)$ results from a simple integration along the $z$
axis. For the deformation potential $F_\alpha^{DP}(x,y) = \nabla^2 \sum_{m,n} A_{mn}^{\alpha} \, X_m(x)
\, Y_n(y)$, whereas for the piezoelectric interaction $F_\alpha^{PZ}(x,y) = \frac{\partial^2}{\partial
x \,\partial y} \sum_{m,n} A_{mn}^{\alpha} \, X_m(x) \, Y_n(y)$. Notice that the Laplacian is a second
order operator, therefore the function $F_\alpha^{DP}$ has the same parity as the phonon mode $\alpha$.
On the other hand, $F_\alpha^{PZ}$ results from first order derivatives, so it has the opposite parity
of $\alpha$.

\begin{figure}[h]
\includegraphics[width=8.6cm]{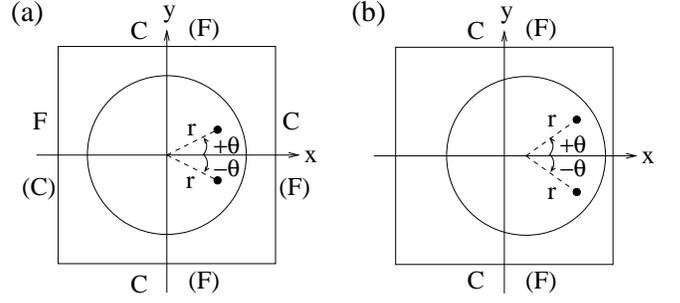}
\caption{Circular quantum dot positioned at: (a) locus A for the \{FCCC\} (or \{CFFF\}) plate; and (b)
locus D for the \{$X_1X_2$CC\} (or \{$X_1X_2$FF\}) plate. Notice that both cases have overall symmetry
only about the $x$ axis, {\it i.e.}, S$_x$.} \label{localization}
\end{figure}

A first examination of Eq. (\ref{integralcircle}) reveals that if $l'' =  l'$ the sine function
vanishes and the matrix element is real. For $l'' \neq  l'$, it will be complex, real or purely
imaginary depending on the system's characteristics. For the sake of understanding we shall briefly
analyze a representative case. Let us assume the same boundary conditions for $Y_1$ and $Y_2$, then
consider two scenarios: the circular QD at locus A and different boundary conditions for $X_1$ and
$X_2$ (Fig. \ref{localization}(a)), or locus D regardless of $X_1$ and $X_2$ (Fig
\ref{localization}(b)). In both cases only the phonon parity about the S$_x$ axis will be relevant for
the evaluation of (\ref{integralcircle}). In  fact, the $+$ ($-$) parity of $F_\alpha$ about S$_x$
leads to a matrix element that is real (purely imaginary) for the DP interaction and purely imaginary
(real) for the PZ interaction. That is a consequence of the parity of the sine and cosine functions
about $\theta=0$ together with the parity of $F_\alpha$ regarding the same axis.

On the basis of the above analysis and the discussion of Section \ref{va}, it follows that if there is
one or more global symmetry axes in the system, then each matrix block ${\bf n} \times {\bf n}$ of Fig.
\ref{blockstructure}(a) can be written as $\mathbb{A} + i \mathbb{B}$, with $\mathbb{A}$ and
$\mathbb{B}$ originating from the cosine and sine parts of the integral in Eq. (\ref{integralcircle}),
respectively. Moreover, those matrices are  real symmetric and mutually disjoint, that is, for
$\mathbb{A}_{r s} \neq 0$ ($\mathbb{B}_{r s} \neq 0$) necessarily $\mathbb{B}_{r s} = 0$
($\mathbb{A}_{r s} = 0$).

Therefore, when a single interaction mechanisms is acting, we have the following scenarios:

\begin{itemize}
\item If the geometrical configuration of the nanostructure is such that there is only one global
symmetry axis (e.g., locus A or D for plate \{CFFF\}), the ensuing partial symmetry break is enough to
generate chaos. Moreover, the matrix representation of the Hamiltonian $\hat{H} = \hat{H}_0 +
\hat{H}_{DP(PZ)}$ can be written in blocks of fixed $l$'s as $\mathbb{H}_{l''l'} = \mathbb{A}_{l''l'} +
(i)^{\mbox{\scriptsize{sign}}(l'' - l')} \, \mathbb{B}_{l''l'}$, where $\mathbb{A}$ and $\mathbb{B}$
are disjoint, real and symmetric. Thus, $\mathbb{H}$ is completely characterized by orthogonal
matrices, so belonging to the GOE universality class. It is straightforward to verify that the present
reasoning encompasses all the cases of a single GOE statistics listed in Table I.

\item For locus A and \{CCFF\} boundary conditions,  the above structure for $\mathbb{H}$ is still
valid. However, now the system has two symmetry axes, leading to new restrictions for the matrix
elements. In fact, denote by $\sigma_x \sigma_y$ (with $\sigma = \pm$) the $\alpha$ mode parities with
respect to S$_x$ and S$_y$. One finds that the integral over the cosine (sine) in Eq.
(\ref{integralcircle}) is different from zero only if $|l'' -l'|$ is even and the $\alpha$ mode is $++$
($--$), or $|l'' -l'|$ is odd and the $\alpha$ mode is $+-$ ($-+$). Such selection rules produce two
different families of eigenvalues for the problem. For \{CCFF\} (and \{FFCC\}) each distinct family is
chaotic, explaining the occurrence of two superposed uncorrelated Wigner distributions in the $P(s)$
statistics (for an explicit example, see the simpler case of a rectangular QD in Sec. V-D).

\item  For locus A and boundary conditions \{CCCC\} (or \{FFFF\}) there exists one further global
symmetry, namely, the equivalence of the $x$ and $y$ directions. The extra symmetry prevents the
emergence of chaos.

\item Finally, in the absence of a global symmetry axis ({\it e.g.}, the quantum dot at C for any
boundary condition, or loci B, C or E for \{CFFF\}) the Hamiltonian matrix does not separate into real
and purely imaginary disjoint parts. Hence, it is a full complex unitary matrix and the chaotic
behavior takes place with the system belonging to the GUE universality class.
\end{itemize}

The last case to be considered is the inclusion of both interactions in the Hamiltonian. From the
previous discussion we know that for a given parity of the mode $\alpha$ the DP and PZ potentials lead
to exactly opposite types of matrix elements. Indeed, if the DP matrix element is real (pure
imaginary), necessarily that corresponding to PZ is pure imaginary (real). Therefore, the Hamiltonian
$\hat{H} = \hat{H}_0 + \hat{H}_{DP} + \hat{H}_{PZ}$ has a complex matrix representation that results in
a GUE statistics for the energy levels.

\subsection{Rectangular quantum dot}

\begin{figure}[h]
\includegraphics[width=4cm]{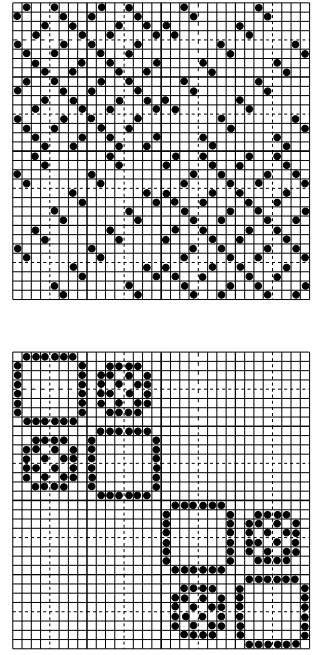}
\caption{ Top: schematics of the $32 \times 32$ interaction matrix for locus D in the \{CCFF\}
nanostructure, ordered as $p \, q \, \alpha_1 \, \alpha_2 \, \alpha_3$ with $p,q = 1 \ \mbox{or} \ 2$
and $\alpha_j = 0 \ \mbox{or} \ 1$. Bottom: the transformed matrix in a block form.  The filled dots
indicate the nonzero elements. } \label{matrixsquare}
\end{figure}

For the rectangular quantum dot the element ${\cal I}_{\alpha \,
\kappa''\kappa'}$ can be written as
\begin{eqnarray}
{\cal I}_{\alpha \, \kappa''\kappa'} &=& f(d)
\int_{\mathcal{D}_{x y}} dx \, dy \, F_\alpha(x,y) \nonumber \\
& & \sin[p''\pi \frac{(x-\overline{x})}{L_x}] \,
\sin[q''\pi \frac{(y-\overline{y})}{L_y}] \nonumber \\
& & \times \sin[p'\pi \frac{(x-\overline{x})}{L_x}] \,
\sin[q'\pi \frac{(y-\overline{y})}{L_y}],
\label{integralsquare}
\end{eqnarray}
where $\overline{x} = x_0 - L_x/2$, $\overline{y} = y_0 - L_y/2$ and $(x_0,y_0)$ are the coordinates of
the center of the QD (for guidance refer to Section \ref{el-gas}). From Eq. (\ref{integralsquare}) it
is evident that the matrix elements are always real numbers. Consequently, any chaotic behavior must
belong to the GOE class and the occurrence of the GUE statistics is ruled out for this nanostructure.

With respect to the quantum numbers, the conditions for which the above integral is different from zero
are again entirely dependent on the global symmetries of the system. For instance, if the whole
nanostructure has S$_x$ as a symmetry axis, then ${\cal I}_{\alpha \, \kappa''\kappa'}$ is nonzero for
the following combinations: $p''+ p'\equiv$ even and $\alpha$ mode parity $\sigma_x \equiv +$, or $p''+
p' \equiv$ odd and $\sigma_x \equiv -$. Similar relations hold for $q''+ q'$ regarding S$_y$.

The behavior of the spectral statistics generated by the rectangular QD-NEMS can be summarized by the
following representative situations: for the loci B, C or E in the \{CFFF\} plate there are no global
symmetry axes and we obtain GOE distributions. For loci A and D in the \{CFFF\} plate, there is one
overall symmetry axis (S$_x$) and the resultant statistics is the superposition of two uncorrelated GOE
distributions. Finally, for locus A in the \{CCFF\} and \{FFFF\} (or \{CCCC\}) plates, which contain
more than one global symmetry axes, no chaotic features are observed. One can verify that all cases in
Table II follow the same trends.

In order to visualize the occurrence of the 2-GOE statistics, consider the case D in the \{CCFF\}
plate. Despite the fact that the phonon modes $\alpha$ have two symmetry axes, S$_x$ and S$_y$, only
the parity about S$_x$ is a global symmetry, due to the position of the QD. Assume then 3 phonon modes,
such that a basis state is written as $|p,q;n_1,n_2,n_3\rangle$, with $p,q$=1 or 2 and $n_{\alpha}$=0
or 1. In addition, the 3 phonon mode parities with respect to S$_x$ are taken to be \{$+,-,+$\}. It
results in the $32 \times 32$ matrix schematically represented in the top of Fig. \ref{matrixsquare},
where the filled dots indicate the nonzero elements. It is possible to transform the original matrix in
that shown at the bottom of Fig. \ref{matrixsquare} just by rearranging its rows and columns. By
labelling the original rows (from left to right) and columns (from top to bottom) as $1,2,\ldots,32$,
we obtain the first nonzero block of the transformed matrix by performing the operation $1 \, 2 \, 3 \,
4 \, 5 \, 6 \, 7 \, 8 \, 9 \, 10 \, 11 \, 12 \, 13 \, 14 \, 15 \, 16 \rightarrow 1 \, 2 \, 5 \, 27 \,
13 \, 19 \, 10 \, 9 \, 32 \, 31 \, 6 \, 20 \, 14 \, 28 \, 23 \, 24$. A similar procedure, {\it i.e.},
operating over the remaining $17 \ldots 32$ positions, leads to the other nonzero block. Here, the
el-ph interaction generates chaos in each family of eigenvalues, originating from the two independent
blocks. Consequently, the spectrum of the full matrix gives rise to the superposition of two
uncorrelated GOE distributions. The above analysis is valid for any matrix size.

Finally, if both interactions act together in a system with a single global symmetry axis, say S$_x$,
their effect is to break the selection rules previously described. This happens because
$F_{\alpha}(x,y)$ has opposite parities for the DP and PZ interactions. As a result the Hamiltonian
matrix does not have a block form and a pure GOE statistics emerges from the 2-GOE case, as seen in
Fig. \ref{DPZ}.

\subsection{Symmetry operator analysis of the anomalous GUE statistics}

So far we have examined the structure of the Hamiltonian matrix to explain the chaotic features
exhibited by our NEMS. Here, we make a link between our results and a more general analysis
\cite{Leyvraz} to clarify the appearance of the anomalous GUE statistics in our time-reversal invariant
(TRI) system.

As already mentioned, the spectral fluctuations of TRI chaotic systems typically correspond to the GOE
distribution. However, Leyvraz et al. \cite{Leyvraz} have shown that there are exceptions to this rule,
which can be interpreted even semiclassically \cite{Keating}. Suppose a TRI chaotic system that has a
discrete point symmetry represented by the operator ${\cal S}$, then $[H,{\cal S}] = [H,{\cal T}] = 0$,
where $H$ is the Hamiltonian and ${\cal T}$ the time reversal operator. More importantly for the
effect, assume also that ${\cal S}$ has two invariant subspaces whose representations are complex
conjugate of each other. We call them $\{\Psi^{(+)}\}$ and $\{\Psi^{(-)}\}$, which are solutions of $H
\Psi_n^{(\pm)} = E_n^{(\pm)} \Psi_n^{(\pm)}$. Since ${\cal T} \Psi_n^{(\pm)} = [{\Psi_n^{(\pm)}}]^* =
\Psi_n^{(\mp)}$, it may seem that the problem is not TRI because each subspace changes into the other
under ${\cal T}$, therefore causing a GUE statistics (notice that the Hamiltonian matrix is complex
Hermitian in this basis). However, this is just an artefact of the particular structure of the
subspaces. Actually, the full Hilbert space is TRI, as can be verified after the simple basis
transformation $\Phi^{(\pm)} = i^{(\pm 1 - 1)/2} [\Psi^{(+)} \pm \Psi^{(-)}]/\sqrt{2}$, for which
${\cal T} \Phi_n^{(\pm)} = {\Phi_n^{(\pm)}}$. Note also that the Kramers theorem
\cite{Sakurai,Dembowski2} imposes $E_n^{(+)} = E_n^{(-)}$. Finally, as pointed out in Ref.
[\onlinecite{Leyvraz}], the present phenomenon is rare because often there exists an extra operator
${\cal P}$ (e.g., the parity symmetry operator) for which $[H,{\cal P}] = [{\cal T},{\cal P}] =0$. This
operator is responsible for combining the two complex conjugate representations of ${\cal S}$ into an
irreducible representation that is self-conjugate \cite{Schafer}, therefore producing a GOE statistics.

Prior to our earliest paper \cite{rego}, the only systems known to show such behavior were billiards
with three-fold but no mirror (parity) symmetries, which have been realized experimentally in microwave
cavities \cite{Schafer,Dembowski1,Dembowski2}. They are chaotic by construction (due to their
particular geometry) and have their eigenstates composed by complex degenerate doublets (of GUE
statistics) and real singlets (of GOE statistics). It is possible, however, to establish a parallel
between our circular QD-NEMS and these billiards. In our case the electron states, Eq.
(\ref{pureelectron}), naturally provide the necessary complex representation through the angular
momentum quantum number $l$. They are divided in singlets, for $l=0$, and degenerate doublets, for
$l=\pm 1, \pm 2, \ldots$. Of course, the original electron states as well as the phonon states are
regular, but the el-ph coupling generates chaos. If, nevertheless, the boundary conditions and the
location of the QD are such to give rise to an global symmetry axis, the energy-level statistics is of
the GOE type due to the ensuing definite parity. On the other hand, in the absence of an overall
symmetry axis (e.g., location C for any plate), no ${\cal P}$ operator exists and GUE statistics
arises.

As a last comment, we recall that in our system the original electron degeneracies are destroyed by the
interaction with the phonons. Nonetheless, the last behave as a perturbation for the electronic
spectrum, because the energies of the electrons are much higher than those for an individual phonon. It
is important to mention, however, that the occurrence of the doublets is not necessary for the
manifestation of the GUE statistics. Actually, even when an additional small perturbation breaks that
degeneracy, the GUE statistics also arises for each split family of eigenstates. It has been confirmed
experimentally by the study of imperfect three-fold microwave triangular billiards \cite{Dembowski1}.

\section{Conclusion}

We have presented a theoretical study of the electron-phonon coupling in nanoelectromechanical systems
(NEMS) comprised of a suspended dielectric plate and a quantum dot on its surface. It is shown that a
quantum chaotic behavior develops as a result of the el-ph interaction, for a wide range of geometrical
and material parameters of the QD-NEMS. A method is developed to treat this novel class of systems. It
associates the phonons with the vibrational modes of a suspended rectangular plate, for clamped and
free boundary conditions. The electrons are confined to a large QD, of either circular or rectangular
symmetry, and described by the free electron gas approximation. The deformation potential and
piezoelectric interactions are included non-perturbatively in the model, by calculating the
eigenenergies of the NEMS on the basis of the el-ph states.

By performing standard energy-level statistics we demonstrate that the resulting spectral fluctuations
are very well described by those of the Gaussian Orthogonal Ensemble (GOE) or the Gaussian Unitary
Ensemble (GUE). It is evidenced that the combination of the phonon mode parities together with the
position of the QD determine the overall symmetries of the system, which are ultimately the responsible
for the distinct chaotic features observed. Although, quantum chaos is commonly obtained in the system,
the GUE statistics occurs only in the case of a circular QD-NEMS. It represents an anomalous
phenomenon, since the problem is time-reversal invariant. The fundamental reason for this effect lies
in the structure of the electronic spectrum, which is formed by doublets with $l=\pm 1, \pm 2, \ldots$.
In the absence of any overall geometrical symmetry, the complex conjugate doublets transform into each
other under the action of the time reversal operator, thus simulating the behavior of a non-TRI system.

Finally, calculations are under way to include the effects of the electron-electron interaction in the
model. We conjecture that the same chaotic behavior can also arise in this case, because the el-el
interaction preserves the total angular momentum of the electronic system, justifying the previous
analysis.

%%%%%%%%%%%%%%%%%%%%%%%%%%%%%%%%%%%%%%%%%%%%%%%%%%
\section*{Acknowledgments}
%%%%%%%%%%%%%%%%%%%%%%%%%%%%%%%%%%%%%%%%%%%%%%%%%%

We thank CNPq/Edital Universal, Funda\c c\~ao Arauc\'aria,
Finep/CT-Infra1, CNPq/CT-Energ and CNPq (MGEL and AG) for research
grants.


\begin{thebibliography}{99}
\bibitem{orthonormal}
To achieve the orthonormalization condition for the one-dimensional
modes with $m > 10$ the Multiprecision package MPFUN was used to
calculate the wave vectors $k_m$ with a 200 digit precision.
\end{thebibliography}
\end{document}